\documentclass[12pt]{article}
\usepackage{epsfig,amsfonts}
\begin{document}

\voffset -0.7 true cm
\hoffset 1.1 true cm
\topmargin 0.0in
\evensidemargin 0.0in
\oddsidemargin 0.0in
\textheight 8.6in
\textwidth 7.1in
\parskip 10 pt

\newcommand{\be}{\begin{equation}}
\newcommand{\ee}{\end{equation}}
\newcommand{\bea}{\begin{eqnarray}}
\newcommand{\eea}{\end{eqnarray}}
\newcommand{\beas}{\begin{eqnarray*}}
\newcommand{\eeas}{\end{eqnarray*}}

\def\kl{{\frac{2 \pi l}{\beta}}}
\def\km{{\frac{2 \pi m}{\beta}}}
\def\kn{{\frac{2 \pi n}{\beta}}}
\def\kr{{\frac{2 \pi r}{\beta}}}
\def\ks{{\frac{2 \pi s}{\beta}}}
\def\b{{\beta}}
\font\cmsss=cmss8
\def\C{{\hbox{\cmsss C}}}
\font\cmss=cmss10
\def\bigC{{\hbox{\cmss C}}}
\def\scriptlap{{\kern1pt\vbox{\hrule height 0.8pt\hbox{\vrule width 0.8pt
  \hskip2pt\vbox{\vskip 4pt}\hskip 2pt\vrule width 0.4pt}\hrule height 0.4pt}
  \kern1pt}}

\begin{titlepage}
\begin{flushright}
{\small BROWN-HET-1264} \\
{\small CU-TP-1017} \\
{\small hep-th/0105171}
\end{flushright}

\begin{center}

\vspace{2mm}

{\Large \bf Black Hole Entropy from Non-Perturbative Gauge Theory}

\vspace{3mm}

Daniel Kabat${}^{1}$, Gilad Lifschytz${}^{2,3}$\ and David A.\ Lowe${}^4$

\vspace{1mm}

${}^1${\small \sl Department of Physics} \\
{\small \sl Columbia University, New York, NY 10027} \\
{\small \tt kabat@phys.columbia.edu}
\vspace{1mm}

${}^2${\small \sl Department of Physics} \\
{\small \sl Princeton University, Princeton, NJ 08544}
\vspace{1mm}

${}^3${\small \sl Department of Mathematics and Physics} \\
{\small \sl University of Haifa at Oranim, Tivon 36006, ISRAEL} 
{\small \tt giladl@research.haifa.ac.il}

${}^4${\small \sl Department of Physics} \\
{\small \sl Brown University, Providence, RI 02912} \\
{\small \tt lowe@het.brown.edu}

\end{center}

\vskip 0.3 cm

\noindent
We present the details of a mean-field approximation scheme for the
quantum mechanics of $N$ D0-branes at finite temperature.  The
approximation can be applied at strong 't Hooft coupling.  We find
that the resulting entropy is in good agreement with the
Bekenstein-Hawking entropy of a ten-dimensional non-extremal black
hole with 0-brane charge.  This result is in accord with the duality
conjectured by Itzhaki, Maldacena, Sonnenschein and Yankielowicz.  We
study the spectrum of single-string excitations within the quantum
mechanics, and find evidence for a clear separation between light and
heavy degrees of freedom.  We also present a way of identifying the
black hole horizon.

\end{titlepage}

\section{Introduction}

The physics of black holes has played a prominent role in our quest to
understand quantum gravity.  Semiclassical considerations have shown
that the horizon of a black hole has an associated thermodynamic
entropy \cite{Hawking}, and a key test of any proposed theory of
quantum gravity should be to provide a microscopic explanation of this
entropy.

Dramatic progress was made a few years ago, when certain extremal
black holes were realized as collections of D-branes in string theory.
This description led to a precise counting of microstates, which was
in exact agreement with semiclassical black hole thermodynamics
\cite{StromingerVafa}.  Unfortunately this counting relied on
supersymmetric non-renormalization theorems, and therefore could only
be applied to certain classes of extremal black holes.

A more general understanding of black hole entropy requires a
non-perturbative definition of string theory.  This is now available,
at least in certain backgrounds, thanks to the M(atrix) and Maldacena
conjectures \cite{BFSS, Maldacena} (for reviews see \cite{reviews}).
These conjectures relate non-perturbative string theories to dual
strongly-coupled large-$N$ gauge theories.  In this framework, black
hole entropy is identified with the entropy of the density matrix
which describes the gauge theory at finite temperature.

In principle, one can use these dualities to understand black hole
physics in terms of gauge theory dynamics.  In practice, however, this
requires two things: a precise map between gravity and gauge theory
quantities, and a tractable calculational scheme in the gauge theory.
Some progress has been made on the first issue \cite{bulk}, 
although even such basic properties as spacetime
locality are still obscure from the gauge theory point of view. 

In this paper we will focus on the second issue, of developing
practical methods for doing gauge theory calculations.  The gauge
theory is strongly coupled whenever semiclassical gravity is valid, so
we must study the gauge theory non-perturbatively.\footnote{In
M(atrix) theory, one can argue that the entropy of certain black holes
is not renormalized beyond one loop \cite{LoweI}.  The argument
assumes Lorentz invariance is recovered in the large $N$ limit, a
property which has been checked at leading order in \cite{LoweII}.}
We do this using techniques from self-consistent mean field theory.
This provides us with an approximation to the density matrix which
describes the gauge theory at finite temperature.  A key test of our
approximation is whether it reproduces the semiclassical
thermodynamics of the black hole.  As we will see, according to this
criterion our approximation works quite well, at least over a certain
range of temperatures.

For simplicity we will concentrate on the quantum mechanics of $N$
D0-branes, with sixteen supercharges and gauge group $SU(N)$
\cite{ClaudsonHalpern}.  At large $N$ and finite temperature, the
effective 't Hooft coupling of the quantum mechanics is
\be
\label{EffectiveCoupling}
g_{\rm eff}^2 = g^2_{YM} N / T^3\,.
\ee
Note that the quantum mechanics is strongly coupled at low
temperature.  This quantum mechanics is dual to a ten-dimensional
non-extremal black hole in type IIA supergravity, with N units of
0-brane charge \cite{imsy}.  The metric of the black hole is
\bea
\nonumber
ds^2 & = & \alpha'\left[-h(U)dt^2 + h^{-1}(U)dU^2 + {c^{1/2}
(g_{YM}^2 N)^{1/2} \over U^{3/2}} d\Omega_{8}^{2}\right] \\
\label{metric}
h(U) & = & \frac{U^{7/2}}{c^{1/2} (g_{YM}^2 N)^{1/2}}\left(1-\frac{U_{0}^{7}}{U^{7}}\right)
\eea
where $c = 2^7 \pi^{9/2} \Gamma(7/2)$ and $g_{YM}$ is the Yang-Mills
coupling constant.  The horizon of the black hole is at $U=U_0$, which
corresponds to a Hawking temperature
\be
\label{HawkingTemp}
T = {7\over 2\pi \sqrt{30}}~ (g^2_{YM} N)^{-1/2} \left(U_0 \over 2 \pi \right)^{5/2} = 0.2034 ~ (g^2_{YM} N)^{-1/2} \left(U_0 \over 2 \pi \right)^{5/2}\,.
\ee
The dual quantum mechanics is to be taken at the same finite
temperature.  The black hole has a free energy, which arises from its
Bekenstein-Hawking entropy \cite{KlebanovTseytlin}.
\begin{equation}
\label{beken}
\beta F = - \left( { 2^{21} 3^2 5^7 \pi^{14} \over
7^{19}}\right)^{1/5} \!\! N^2 \left({T \over (g_{YM}^2 N)^{1/3}}
\right)^{9/5} = - 4.115 ~ N^2 \left({T \over (g_{YM}^2 N)^{1/3}} \right)^{1.8}
\end{equation}
Duality predicts that the quantum mechanics should have the same free
energy.  The supergravity description is expected to be valid when the
curvature and the dilaton are small near the black hole horizon. This
regime corresponds to the 't Hooft large-$N$ limit of the quantum
mechanics, when the temperature is such that the dimensionless
effective coupling (\ref{EffectiveCoupling}) lies in the
range \cite{imsy}
\be
\label{CouplingRange}
1 \ll g_{\rm eff}^2 \ll N^{10/7}\,.
\ee

An outline of this paper is as follows.  In section 2 we develop a
mean-field approximation scheme for 0-brane quantum mechanics,
building on our earlier work \cite{KabatLifschytz}.  In section 3 we
present numerical results for the behavior of the gauge theory,
focusing on thermodynamic quantities.  We compare our results to the
black hole predictions, and find good agreement over a certain range
of temperatures.  Section 4 is devoted to a spectral analysis of
the propagators, to extract the spectrum of stretched strings that
make up the supergravity background. In section 5 we discuss how local spacetime
physics, such as the size of the black hole horizon, may be extracted
from the gauge theory.  Section 6 gives our conclusions
and a discussion of possible future directions.  A summary of our
results has appeared in \cite{KabatLifschytzLowe}.

\section{Mean-field approximation for 0-brane quantum mechanics}

The basic idea of our approximation is to treat the ${\cal O}(N^2)$
degrees of freedom appearing in 0-brane quantum mechanics as
statistically independent, with interactions taken into account via a
sort of mean-field approximation.  In the rest of this section we
present several reasons to believe this simple approximation captures
some of the essential physics of the quantum mechanics in the
supergravity regime.  In the next section we will show that the
approximation gives results which are in good agreement with black
hole thermodynamics over a certain temperature range.

Let us begin by stating our approach to studying strongly-coupled
systems in rather general terms.  We are presented with a
strongly-coupled action $S$, in our case the action for 0-brane
quantum mechanics.  We approximate this action with a simpler trial
action $S_0$.  All quantities of interest can then be computed as an
expansion in powers of $S - S_0$.  For instance, the free energy has
an expansion \cite{Feynman}
\bea
\label{betaFexpansion}
\beta F & = & \beta F_0 - \langle e^{- (S - S_0)} - 1 \rangle_{\C,0} \\
\nonumber
& = & \beta F_0 + \langle  S - S_0 \rangle_0
- {1 \over 2} \langle  \left(S - S_0\right)^2\rangle_{\C,0} + \cdots
\eea
where a subscript $\bigC,0$ denotes a connected expectation value
calculated using the trial action $S_0$.  If the trial action comes
sufficiently close to capturing the dynamics of the full action, then
this expansion should be well-behaved, even if the full action $S$ is
strongly coupled.

This sort of approximation relies crucially on an appropriate choice
of trial action.  In our case, we shall take $S_0$ to be the most
general quadratic action that one can write in terms of the
fundamental gauge theory degrees of freedom.  This means that our
trial action involves an infinite number of adjustable parameters,
namely the momentum-dependent two-point functions of all the
fundamental fields.  One can regard these propagators as providing an
infinite set of variational parameters.  To fix these parameters we
solve a truncated set of Schwinger-Dyson equations.  These gap equations
provide a non-perturbative approximation to the true two-point
functions of the theory, by resumming an infinite set of Feynman
diagrams.

As we shall see, this sort of approximation has several attractive
features, which initially motivated us to apply these techniques to
0-brane quantum mechanics.
\begin{itemize}
\item The approximation is non-perturbative in the Yang-Mills coupling
constant, and self-consistently cures the infrared divergences which
are present in conventional finite-temperature perturbation theory.
This makes it possible to apply the approximation at strong coupling,
at temperatures where one can make a direct comparison with black hole
predictions.
\item We can formulate the approximation in a way which respects 't
Hooft large-$N$ counting, by only keeping planar contributions to the
Schwinger-Dyson equations.  This means that an overall factor of $N^2$
in the free energy, as well as the appearance of the gauge coupling
only in the combination $g^2_{YM} N$, is guaranteed.  But this is
exactly the form (\ref{beken}) of the supergravity result!  That is,
we are proposing that the overall factor of $N^2$ in the supergravity
free energy can be understood in terms of ${\cal O}(N^2)$ elementary
quasiparticles, which are in one-to-one correspondence with the
degrees of freedom appearing in the fundamental Lagrangian.\footnote{A
``Fermi liquid'' approach to black hole physics!}  Incidentally, this
means that our approximations are hopeless at couplings (outside the
range (\ref{CouplingRange})) which are so strong that 't Hooft scaling
breaks down.
\item A key feature of the approximation is that a quadratic trial
action will automatically respect all symmetries that act linearly on
the fundamental fields.  This is crucial in a problem like 0-brane
quantum mechanics, where symmetries play such an important role.  By
working in a superfield formalism with off-shell supersymmetry, our
trial action will have ${\cal N} = 2$ supersymmetry and $SO(2) \times
SO(7)$ rotational symmetry (out of the underlying ${\cal N} = 16$
supersymmetry and $SO(9)$ rotational symmetry).
\item Another feature of the approximation is that it avoids
certain infrared problems which are present in the full 0-brane
quantum mechanics.  The difficulty is that the partition function of
the full quantum mechanics contains an infrared divergence from the
regions in moduli space where the 0-branes are far apart. This leads
to a divergent contribution to the entropy with an overall coefficient
${\cal O}(N)$.  From the supergravity point of view, this corresponds
to a thermal gas of gravitons. This divergence may be regulated by
putting the system in a finite box. The black hole entropy which is
${\cal O}(N^2)$ can then easily be made to dominate over the ${\cal
O}(N)$ contribution.  Our mean-field approximation automatically
computes the ${\cal O}(N^2)$ piece, while discarding the subleading
${\cal O}(N)$ divergence, so no additional infrared regularization is
required.
\end{itemize}
This sort of approximation also has some potential drawbacks.
\begin{itemize}
\item An unfortunate fact is that there is no {\em a priori} guarantee
that the approximation works well.  One has to choose a trial action
and a set of gap equations, and hope that with appropriate choices the
approximation works well.  In our case, we will be able to justify our
choices {\em a posteriori} by showing that we get good agreement with
black hole thermodynamics over a certain temperature range.  Another
way to justify the approximation is to compute higher-order terms in
the expansion (\ref{betaFexpansion}) and show that they are small.  We
have not attempted this for the full 0-brane problem, although toy
models show promising behavior \cite{KabatLifschytz}.
\item Although the approximation respects all symmetries which act
linearly on the fields, it breaks symmetries that act non-linearly.
As there is no superspace formulation of theories with 16
supercharges, we can only realize a subgroup of the supersymmetries
(in our case ${\cal N} = 2$) as acting linearly on the fields.  This
is sufficient, for example, to make our approximation to the
vacuum energy vanish as $\beta \rightarrow \infty$.  However, the
remaining supersymmetries and R-symmetries are broken by the
approximation.  Another important symmetry which acts non-linearly on
fields, and is therefore broken by our approximation, is gauge
invariance.  More precisely, our quadratic trial action is not
invariant under BRST transformations.  As we shall show in section
2.3, this difficulty can be largely overcome by an appropriate gauge
choice.
\end{itemize}

Before presenting the details of the approximation, let us note that
the techniques we are using have a long history.  They are closely
related to variational methods \cite{Feynman} and self-consistent
Hartree-Fock approximations, and also go by the name of modified
perturbation theory \cite{Okopinska}.  They are equivalent to the
effective action formalism developed in \cite{CJT}.  Similar
techniques have been applied to QCD \cite{QCD}, and related techniques
are used to study finite-temperature field theory \cite{DolanJackiw}.
Our own work on the subject began with \cite{KabatLifschytz}, where
we were motivated by the 0-brane problem to apply these techniques to
several toy problems in supersymmetric quantum mechanics.  Related
techniques have been applied to $0+0$ dimensional Yang-Mills integrals
in \cite{OdaSugino}, and have also been used to study Wilson loops in
${\cal N}=4$ gauge theory in \cite{EricksonSemenoffZarembo}.

\subsection{The 0-brane action in ${\cal N} = 2$ superspace}

We begin by formulating the 0-brane action in ${\cal N} = 2$
superspace.  For more details see appendix A of \cite{KabatLifschytz}.

${\cal N} = 2$ supersymmetry means that we have an $SO(2)$ R-symmetry,
with spinor indices $\alpha,\beta = 1,2$ and vector indices $i,j =
1,2$.  The $SO(2)_R$ Dirac matrices $\gamma^i_{\alpha\beta}$ are real,
symmetric, and traceless.  Given two spinors $\psi$ and $\chi$, there
are two invariants one can make, which we denote by
\[
\psi_\alpha \chi_\alpha \qquad {\rm and} \qquad \psi^\alpha
\chi_\alpha \equiv {i \over 2} \epsilon_{\alpha\beta}
\psi_\alpha \chi_\beta\,.
\]
${\cal N} = 2$ superspace has coordinates $(t,\theta_\alpha)$, where
$\theta_\alpha$ is a collection of real Grassmann variables that
transform as a spinor of $SO(2)_R$.  The simplest representation of
supersymmetry is a real scalar superfield
\[
\Phi = \phi + i \psi_\alpha \theta_\alpha + f \theta^2\,.
\]
It contains a physical real boson $\phi$ and a physical real fermion
$\psi_\alpha$, along with a real auxiliary field $f$.  To describe
gauge theory we introduce a real spinor connection on superspace
$\Gamma_{\alpha}$, with component expansion
\[
\Gamma_\alpha = \chi_\alpha + A_0 \theta_\alpha + X^i
\gamma^i_{\alpha\beta} \theta_\beta + d \epsilon_{\alpha\beta}
\theta_\beta + 2 \epsilon_{\alpha\beta} \lambda_\beta \theta^2 \,.
\]
The fields $X^i$ are physical scalars, while $\lambda_\alpha$ are
their superpartners, $d$ is an auxiliary boson, $\chi_\alpha$ are
auxiliary fermions, and $A_0$ is the 0+1 dimensional gauge field.

To write a Lagrangian we introduce a supercovariant derivative
\be
D_\alpha  =  {\partial \over \partial \theta_\alpha} - i \theta_\alpha
{\partial \over \partial t}
\ee
and its gauge-covariant extension
\be
\nabla_\alpha = D_\alpha + \Gamma_\alpha\,.
\ee
The 0-brane action is built from a collection of seven adjoint scalar
multiplets $\Phi_a$ that transform in the ${\bf 7}$ of a $G_2 \subset
SO(9)$ global symmetry, coupled to a $U(N)$ gauge multiplet
$\Gamma_\alpha$.  The action reads
\be
\label{action}
S_{SYM}  =  {1 \over g^2_{YM}} \int dt d^2\theta \, {\rm Tr} \biggl(
- {1 \over 4} \nabla^\alpha {\cal F}_i \nabla_\alpha {\cal F}_i
- \frac{1}{2} \nabla^\alpha \Phi_a \nabla_\alpha \Phi_a
- \frac{i}{3} f_{abc} \Phi_a [\Phi_b, \Phi_c] \biggr)\,.
\ee
Here ${\cal F}_i = {1 \over 4} \gamma^i_{\alpha\beta} \lbrace
\nabla_\alpha, \nabla_\beta \rbrace$ is the field strength constructed
from $\Gamma_\alpha$, and $f_{abc}$ is a totally
antisymmetric $G_{2}$-invariant tensor, normalized to
satisfy\footnote{This corrects a normalization error in
\cite{KabatLifschytz}.}
\be
f_{abc} f_{abd} = {3 \over 2} \delta_{cd}\,.
\ee
Strictly speaking, the relative positions of the $N$ 0-branes are
governed by an $SU(N)/{\mathbb Z}_N$ gauge theory, but in the
large-$N$ limit we can approximate this by $U(N)$.

We are interested in the finite temperature properties of the action
(\ref{action}).  We work in Euclidean space, setting
\[
S_E = -i S_M, \qquad \tau = i t, \qquad A_0{}_E = -i A_0{}_M, \qquad f_E = -i f_M\,.
\]
Note that we must Wick-rotate the auxiliary fields, to get a Euclidean
action that is bounded below.  As usual we compactify the Euclidean
time direction on a circle of circumference $\beta$, which is identified with
the inverse temperature.  Bosons are periodic while fermions are
antiperiodic; for example we write the mode expansions
\beas
X^i(\tau) & = & {1 \over \sqrt{\beta}} \sum_{l \in {\mathbb Z}} X^i_l\,
e^{i 2 \pi l \tau / \beta} \\
\chi_\alpha(\tau) & = & {1 \over \sqrt{\beta}} \sum_{r \in {\mathbb Z} + 1/2}
\chi_{\alpha\,r}\, e^{i 2 \pi r \tau / \beta}\,.
\eeas

\subsection{Gauge fixing}

Our approximation is based on resumming an infinite class of Feynman
diagrams to obtain an approximation for the two-point functions at
strong coupling.  To make this procedure well-defined, we must fix a
choice of gauge.  For reasons we will explain, it is extremely
advantageous to work in the gauge
\be
\label{GaugeChoice}
D^{\alpha} \Gamma_{\alpha}=0\,.
\ee
The first advantage of this gauge is that, since (\ref{GaugeChoice})
is a condition on superfields, our gauge choice preserves manifest
supersymmetry.  In terms of component fields, it sets
\be
\partial_t A_0 = 0, \qquad
d = 0, \qquad
\lambda_\alpha = {1 \over 2} \partial_t \chi_\alpha\,.
\ee
This is a complete gauge fixing, {\em i.e.}~having made this choice
there is no residual freedom to make additional gauge transformations.

A second advantage is that our gauge choice is well-defined at finite
temperature.  To see this, note that the zero mode of the gauge field,
which we denote $A_{00}$, survives as a physical degree of freedom.
This is important, because at finite temperature the gauge theory
acquires an additional dynamical degree of freedom, namely the value of
a Wilson-Polyakov loop around the Euclidean time direction $U = P e^{i
\oint d\tau A_0}$.  In our gauge, this physical degree of freedom is
parameterized by the zero mode.
\be
U  = e^{i \sqrt{\beta} A_{00}}
\ee
Note that at finite temperature
\be
\label{AooPeriodicity}
A_{00} \sim A_{00} + 2 \pi/\sqrt{\beta}
\ee
is a periodic variable.

Corresponding to our choice of gauge we must introduce a ghost action
(but no gauge fixing term)
\[
S_{ghost} = {1 \over g^2_{YM}} \int dt d^2\theta \, {\rm Tr} \left(
D^\alpha \bar{C} \nabla_\alpha C \right)\,.
\]
For the ghost multiplet we adopt the component expansion
\[
C = \alpha + \beta_\alpha \theta_\alpha + \gamma \theta^2
\]
where $\alpha$ and $\gamma$ are complex Grassmann fields and
$\beta_\alpha$ is a complex boson.  At finite temperature $\alpha$ and
$\gamma$ are periodic, while $\beta$ is antiperiodic.

\subsection{Slavnov-Taylor identities}

Mean-field methods usually have a difficult time dealing with gauge
symmetry.  The problem is that the Slavnov-Taylor identities
are typically violated by the approximation.  After gauge fixing, Slavnov-Taylor
identities arise from BRST invariance of the gauge fixed action.  BRST
transformations act non-linearly on fields, but the sort of
mean-field approximation we wish to use is based on a trial action
that is quadratic in the fundamental fields.  Such a trial action
cannot respect a symmetry that acts non-linearly.  Thus mean-field
techniques typically break BRST invariance, and hence violate Slavnov-Taylor
identities.

A major advantage of our gauge choice (\ref{GaugeChoice}) is that many
of the Slavnov-Taylor identities become trivially satisfied, so that even a
quadratic trial action can respect many of the consequences of gauge
invariance.  To illustrate this, we consider a simplified model, which
can be obtained from the full 0-brane quantum mechanics by discarding
all fermion and auxiliary fields.  That is, we study bosonic
Yang-Mills quantum mechanics, with the following gauge-fixed Euclidean
action.
\goodbreak
\bea
\nonumber
S & = & {1 \over g^2_{YM}} \int d\tau \, {\rm Tr} \biggl\lbrace {1 \over 2} D_\tau X^i D_\tau X^i
- {1 \over 4} [X^i,X^j] [X^i,X^j] \\
\label{BosonicAction}
&  & \qquad \qquad \qquad + {1 \over 2 \xi} \left(\partial_\tau A_0\right)^2
+ \partial_\tau \bar{\alpha} D_\tau \alpha \biggr\rbrace
\eea
Here $A_0$ is a $U(N)$ gauge field, with $D_\tau = \partial_\tau + i
[A_0, \cdot]$.  The fields $X^i$ are adjoint scalars, and $\alpha$ is
a ghost field.  One subtle point is that the antighost zero mode
$\bar{\alpha}_{l = 0}$ does not appear in the action, and therefore
should not be regarded as a true degree of freedom.  It is completely
decoupled, and any correlators involving $\bar{\alpha}_0$ vanish.

To illustrate the difficulties with gauge invariance we have adopted a
general class of gauges parameterized by $\xi$. The action is obtained
by gauge fixing $\partial_\tau A_0= f$ and
then functionally integrating over $f$, with the weight
$\exp(-\int d\tau f^2/2 g^2_{YM} \xi)$. Our preferred gauge 
condition $\partial_\tau A_0 = 0$ is recovered in the limit $\xi
\rightarrow 0$.  Expanding the fields in Fourier modes, the action
(\ref{BosonicAction}) is invariant under BRST transformations
\bea
\nonumber
\delta_\eta A_{0\,l} & = & - \eta \left(\kl \alpha_l +{1 \over \sqrt{\beta}} \sum_{m+n = l} [A_{0\,m}, \alpha_n]\right) \\
\nonumber
\delta_\eta X^i_l & = & - \eta {1 \over \sqrt{\beta}} \sum_{m+n = l} [X^i_m, \alpha_n] \\
\nonumber
\delta_\eta \alpha_l & = & \eta {1 \over \sqrt{\beta}} \sum_{m+n = l} \alpha_m \alpha_n \\
\label{BRST}
\delta_\eta \bar{\alpha}_l & = & {\eta \over \xi} \kl A_{0,\,-l}
\eea
where $\eta$ is a Grassmann parameter.  Note that the decoupled
antighost zero mode $\bar{\alpha}_0$ is indeed invariant under BRST
transformations.

We can use this BRST symmetry to derive Slavnov-Taylor identities in the
standard way, from the fact that the expectation value of any
BRST-exact quantity vanishes.  For example, we must have
\be
\label{Ward1}
\langle \delta_\eta \left(\bar{\alpha}_l A_{0\,l}\right)\rangle = 0\,.
\ee
This gives us the following relation among Green's functions.
\be
\label{Ward2}
\langle  {1 \over \xi} \kl A_{0,\,-l} A_{0\,l} \rangle + \langle  \bar{\alpha}_l \biggl(\kl \alpha_l + {1 \over \sqrt{\beta}} \sum_{m+n = l}
[A_{0\,m} \alpha_n]\biggr) \rangle = 0
\ee
For $l = 0$ this Slavnov-Taylor identity is trivially satisfied: the first term
vanishes since $l = 0$, while the second term vanishes since
$\bar{\alpha}_0$ is decoupled. Here we assume the $A_{0,l}$ two-point
function is finite at $l=0$.    
For $l \not= 0$ the second term can be
simplified using the following Schwinger-Dyson equation (a consequence
of the ghost equation of motion)
\be
\left(\kl\right)^2 \langle  {\rm Tr} \left(\bar\alpha_l \alpha_l\right) \rangle
+ {1 \over \sqrt{\beta}} \kl \sum_{m+n = l} \langle  {\rm Tr} \left( \bar{\alpha}_l
[A_{0\,m}, \alpha_n]\right) \rangle = - g^2_{YM} N^2
\ee
where for simplicity we have taken a trace to get rid of matrix
indices.  Thus, modulo the use of an equation of motion, the content
of the identity (\ref{Ward1}) is the well-known fact that the gauge
field propagator at non-zero frequency is given exactly by the
gauge-fixing term in the classical action:
\be
\langle  {\rm Tr} \left( A_{0\,l} A_{0,\,-l}\right) \rangle = {g^2_{YM} N^2 \xi \over (2 \pi l / \beta)^2}
\qquad \hbox{\rm for $l \not= 0$.}
\ee
In the limit $\xi \rightarrow 0$ this Slavnov-Taylor identity implies that the
modes of $A_0$ with non-zero frequency do not propagate.  But this is
an automatic consequence of adopting our gauge choice
(\ref{GaugeChoice}), which eliminates all non-zero modes of $A_0$!  In
fact, in the full 0-brane quantum mechanics, {\em all Slavnov-Taylor identities
which just constrain two-point functions are automatically satisfied
by working in the gauge (\ref{GaugeChoice})}.

Next let us consider a Slavnov-Taylor identity on a 3-point function.  We have the requirement
\be
\label{Ward3}
\langle  \delta_\eta \left( \bar{\alpha}_l X^i_m X^j_n \right) \rangle = 0\,.
\ee
Using the transformations (\ref{BRST}), this gives rise to a Slavnov-Taylor identity with the
schematic form
\be
{1 \over \xi} \langle  \partial_\tau A_0 X^i X^j \rangle = \langle  \bar{\alpha} \alpha X^i X^j \rangle\,.
\ee
If the gauge field carries zero frequency this identity turns out to
be trivially satisfied (for the same reasons that (\ref{Ward2}) was
trivially satisfied at $l = 0$).  If the gauge field carries non-zero
frequency then this Slavnov-Taylor identity is non-trivial.  In particular, in
the limit $\xi \rightarrow 0$, it states that the amplitude to emit a
gauge boson with non-zero frequency is ${\cal O}(\xi)$.  {\em But this
property is automatically satisfied by working in the gauge
(\ref{GaugeChoice}), where the non-zero modes of the gauge field are
eliminated.}  Again, the content of the Slavnov-Taylor identity (\ref{Ward3}) is
automatically taken into account just by working in the gauge
(\ref{GaugeChoice}).

This pattern is quite general.  All non-trivial Slavnov-Taylor identities follow
from the requirement that correlators of the form
\be
\langle  \delta_\eta \left( \bar{\alpha}_l \, \cdots \right) \rangle
\ee
vanish.  (There must be at least one $\bar{\alpha}$, since
$\delta_\eta$ increases the ghost number by one and you need zero
ghost number to have a non-vanishing correlator.)  If $l = 0$ this
Slavnov-Taylor identity is trivially satisfied.  If $l \not= 0$ this Slavnov-Taylor
identity becomes a constraint on a correlators that either involve a
gauge boson with non-zero frequency, or involve an antighost with
non-zero frequency.  Correlators with $A_{0,\,l\not=0}$ must vanish in
the limit $\xi \rightarrow 0$, and this property is guaranteed by
working in the gauge (\ref{GaugeChoice}).  In fact it is not clear to
us whether the Slavnov-Taylor identities have any non-trivial content in the
gauge (\ref{GaugeChoice}).  In principle it seems that they could give
constraints on correlators involving antighosts, but at the level of
2-point and 3-point functions, no constraints arise which are
not already implied by the Schwinger-Dyson equations. 

Does this issue of Slavnov-Taylor identities have any practical importance?
After all, the approximation could work well even though it is not
gauge invariant.  But it turns out that in our case, the gauge choice
(\ref{GaugeChoice}) is crucial.  We have used mean-field methods to
study gauge theories (including (\ref{action}), (\ref{BosonicAction}))
in the more general $R_\xi$ class of gauges, and have found that the
system of one-loop truncated Schwinger-Dyson equations does not have
solutions when the gauge theory is strongly coupled.  We believe this
breakdown can be related to the fact that the violation of Slavnov-Taylor
identities gets worse as the coupling increases.  

In any case, at
least for 0-brane quantum mechanics, this difficulty can be avoided by
working in the gauge (\ref{GaugeChoice}). The vertices that appear in
the gap equations receive no constraints that are not already implied by
the Schwinger-Dyson equations
(quartic vertices that appear in the gap equations will not involve a
pair of ghosts). The Schwinger-Dyson equations themselves will be
satisfied at one-loop level, so the approximation is self-consistent.

\subsection{Trial action and gap equations}

In applying mean-field methods to 0-brane quantum mechanics, the first
step is to choose a trial action.  We will adopt the following trial
action, which is written in terms of component fields expanded in
Matsubara modes.
\bea
\label{TrialAction}
\nonumber
S_0 & = & - {N \over \lambda} {\rm Tr} (U + U^\dagger) + \sum_l \frac{1}{2 \sigma_l^2} {\rm Tr}(X^i_l X^i_{-l}) 
- \sum_r \frac{1}{2 a_r}{\rm Tr}(\chi_{\alpha r} \chi_{\alpha,\,-r}) \\
\nonumber
& & + \sum_l \frac{1}{2 \Delta_l^2}{\rm Tr}(\phi^a_l\phi^a_{-l}) - \sum_r \frac{1}{2g_r}{\rm Tr}(\psi^{a}_{\alpha r} \psi^a_{\alpha,\,-r})
+ \sum_l \frac{1}{2 \epsilon_l^2}{\rm Tr}(f^a_lf^a_{-l}) \\
& & - \sum_{l \neq 0}\frac{1}{s_l} {\rm Tr} (\bar{\alpha}_l \alpha_l) + \sum_r \frac{1}{t_r}{\rm Tr}(\bar{\beta}_{\alpha r} \beta_{\alpha r})
- \sum_l \frac{1}{u_l}{\rm Tr}(\bar{\gamma}_l\gamma_l)
\eea
Recall that $l,m \in {\mathbb Z}$ and $r,s \in {\mathbb Z} + {1 \over
2}$ label Fourier modes, $\alpha,\beta = 1,2$ are $SO(2)_R$ spinor
indices, $i,j = 1,2$ are $SO(2)_R$ vector indices, and $a,b =
1,\ldots,7$ are indices in the {\bf 7} of $G_2$.  The parameters
$\lambda,\,\sigma_l^2,\,\ldots$ can be thought of as variational
parameters, which we will fix by solving a set of one-loop gap
equations.

The action (\ref{TrialAction}) is essentially the most general
Gaussian trial action that is compatible with the linearly-realized
bosonic symmetries of the problem.\footnote{including the ${\mathbb
Z}_2$ $R$-parity symmetry discussed in appendix A of
\cite{KabatLifschytz}.}  Supersymmetry is broken at finite
temperature, so we have not imposed supersymmetry on the action
(\ref{TrialAction}), although as we discuss below supersymmetry gets
incorporated into our approximation in a natural way.

There are a few subtle points to note about this action.  One point is
that, due to the periodicity (\ref{AooPeriodicity}), it is not
appropriate to adopt a Gaussian trial action for $A_{00}$.  Rather we
have adopted the unitary one-plaquette model action \cite{GrossWitten}
\[
S_\scriptlap = - {N \over \lambda} {\rm Tr} (U + U^\dagger)
\]
for the holonomy $U = e^{i \sqrt{\beta} A_{00}}$.  This action
undergoes a large-$N$ phase transition when $\lambda = 2$.  As
discussed in \cite{KabatLifschytz}, such a transition is expected to
separate the perturbative gauge theory regime from the supergravity
regime, presuming couplings to other fields do not turn this into a
smooth cross-over. 
  A second minor point is that, as discussed in section 2.3,
the antighost zero mode is not a physical degree of freedom.  We have
therefore suppressed the terms involving $\bar{\alpha}_0$ in
(\ref{TrialAction}).

Corresponding to the action (\ref{TrialAction}) we have the 2-point correlators
\bea
\nonumber
&&\!\!\!\!\langle A_{00} A_{00}\rangle_0 \equiv \rho_0^2 \qquad \quad \langle X_l^i X_m^j\rangle_0 = \sigma_l^2 \delta^{ij} \delta_{l+m}
\qquad \langle \chi_{\alpha r} \chi_{\beta s}\rangle_0 = a_r \delta_{\alpha\beta} \delta_{r+s} \\
\nonumber
&&\!\!\!\!\langle \phi^a_l \phi^b_m\rangle_0 = \Delta_l^2 \delta^{ab} \delta_{l+m} \quad\!\!\! \langle \psi^a_{\alpha r} \psi^b_{\beta s}\rangle_0 = g_r \delta^{ab} \delta_{\alpha\beta}
\delta_{r+s} \quad\!\!\! \langle f^a_l f^b_m\rangle_0 = \epsilon_l^2 \delta^{ab} \delta_{l+m} \\
\label{propagators}
&&\!\!\!\!\langle \bar{\alpha}_l \alpha_m\rangle_0 = s_l \delta_{lm} \qquad \quad \langle \bar{\beta}_{\alpha r} \beta_{\beta s}\rangle_0 = t_r \delta_{\alpha \beta} \delta_{rs}
\qquad\,\,\, \langle \bar{\gamma}_l \gamma_m\rangle_0 = u_l \delta_{lm}
\eea
where $\langle \cdots\rangle_0$ denotes an expectation value computed using $S_0$,
and where the two-point function of the gauge field zero mode is given by
\be
\label{rho02}
\rho_0^2 = \left\lbrace
\begin{array}{ll}
{2 \over \beta N} \left[{\rm li}_2\left(1 - {\lambda \over 2}\right)
+ \left(1 - {2 \over \lambda}\right) \log \left(1 - {\lambda \over 2}\right) - 1 \right]
& \lambda \leq 2 \\
\noalign{\vskip 2mm}
{1 \over \beta N} \left({\pi^2 \over 3} - {4 \over \lambda}\right) & \lambda \geq 2
\end{array} \right.
\ee
involving a dilogarithm \cite{KabatLifschytz}.

Next we need to choose a set of gap equations to fix the parameters
that appear in our trial action.  For most degrees of freedom we will
adopt the one-loop gap equations discussed in \cite{KabatLifschytz}.
These equations can be obtained by demanding that the
quantity\footnote{This quantity can be identified with the two-loop
2PI effective action of Cornwall, Jackiw and Tomboulis \cite{CJT}.}
\be
\label{Ieff}
I_{\rm eff} = \beta F_0 + \langle S_{II} + S_{IV} - S_0\rangle_0 - \frac{1}{2} \langle \left(S_{III}\right)^2\rangle_{\C,0}
\ee
is stationary with respect to arbitrary variations of the 2-point
functions (\ref{propagators}), where $S_{II}$, $S_{III}$, $S_{IV}$
refer to terms in the SYM-plus-ghost action that are quadratic, cubic,
quartic in the fundamental fields.  The explicit expression for this
effective action is given in appendix A.

For the gauge field, however, we use a slightly different gap
equation.  The starting point is the Schwinger-Dyson equation for
$\langle {\rm Tr} U\rangle$, which follows from demanding that
\be
\langle U\rangle = \int dU \, d(\cdots) \,U\,e^{-S}
\ee
is invariant under an infinitesimal change of variables $U \rightarrow
gU$ with $g = 1 + i \omega \in U(N)$.  At leading order this implies 
\be
\label{AooSD1}
\langle  {\rm Tr} U \rangle = - {i \over \sqrt{\beta}} \langle  {\rm Tr} ( U {\delta S \over \delta A_{00}} )\rangle
\ee
where we have dropped higher-order terms in $\delta A_{00}$, coming
from the Campbell-Baker-Hausdorff lemma, which do not contribute at
one loop in the mean field approximation.  
Evaluating (\ref{AooSD1}) to one-loop order gives a
relation between expectation values computed with respect to $S_0$.
\bea
\label{AooSD}
{1 \over N} \langle {\rm Tr} U \rangle_\scriptlap & = & {2 \over \beta} \biggl(- {i \sqrt{\beta} \over N} \langle {\rm Tr} (U A_{00}) \rangle_\scriptlap \biggr) \, \cdot \\
\nonumber
& & \cdot \, {\partial \over \partial \rho_0^2} \left(\langle S_{IV}\rangle_0 - {1 \over 2} \langle \left(S_{III}\right)^2\rangle_{\C,0}\right)
\eea
The relevant one-plaquette expectation values are \cite{GrossWitten}
\bea
\label{WilsonVEV}
{1 \over N} \langle {\rm Tr} U\rangle_\scriptlap & = & \left\lbrace
\begin{array}{ll}
1 - \lambda/4 & \quad \lambda \leq 2 \\
\noalign{\vskip 2mm}
1/\lambda & \quad \lambda \geq 2
\end{array} \right. \\
\noalign{\vskip 3mm}
\nonumber
- {i \sqrt{\beta} \over N} \langle {\rm Tr} (U A_{00}) \rangle_\scriptlap & = & \left\lbrace
\begin{array}{ll}
{1 \over 2}\left(1+{\lambda \over 4}\right) - {1 \over \lambda} \left(1 - {\lambda \over 2}\right)^2 \log \left(1 - {\lambda \over 2}\right)
& \quad \lambda \leq 2 \\
\noalign{\vskip 2mm}
1- {1 \over 2 \lambda} & \quad \lambda \geq 2\,,
\end{array} \right.
\eea
while the expressions for $\langle S_{IV}\rangle_0$ and $- {1 \over 2}
\langle \left(S_{III}\right)^2\rangle_{\C,0}$ are given in appendix $A$.  We adopt (\ref{AooSD})
as the gap equation that fixes the one-plaquette coupling $\lambda$.

Let us pause to note a few important features of this system of gap
equations.  First, in computing expectation values we have kept only
planar contributions.  This means 't Hooft large-$N$ counting is
automatic: the free energy will come with an overall factor of $N^2$,
and the Yang-Mills coupling will only appear in the combination
$g^2_{YM} N$.  We henceforth adopt units which set $g^2_{YM} N = 1$,
by rescaling all dimensionful quantities as in \cite{KabatLifschytz}.
Second, since we have consistently worked to one-loop order while
including all auxiliary fields, these gap equations respect
supersymmetry.  Of course supersymmetry gets broken at finite
temperature, but in the zero-temperature limit these symmetry breaking
effects go away.\footnote{Supersymmetry is unbroken at zero
temperature in this model.}  Thus as $\beta \rightarrow \infty$ the
bosonic and fermionic propagators will be related by supersymmetry
Ward identities, and the vacuum energy will automatically vanish.

To summarize, the parameters appearing in our trial action are fixed by solving the following set of gap equations.
\bea
\nonumber
&& {1 \over N} \langle {\rm Tr} U \rangle_\scriptlap = {2 \over \beta} \biggl( - {i \sqrt{\beta} \over N} \langle {\rm Tr} (U A_{00}) \rangle_\scriptlap \biggr)
\biggl({2 \over \beta} \sum_l \sigma_l^2 + {5 i \over 2 \beta} \sum_r \kr a_r \\
\nonumber
&& \quad + {7 \over \beta} \sum_l \Delta_l^2 - {4 \over \beta} \sum_l (\kl)^2 (\sigma_l^2)^2 - {4 \over \beta} \sum_r (\kr)^4 (a_r)^2
- {14 \over \beta} \sum_l (\kl)^2 (\Delta_l^2)^2 \\
\label{gap1}
&& \quad - {7 \over \beta} \sum_r (g_r)^2 + {1 \over \beta} \sum_l (\kl)^2 (s_l)^2 + {1 \over 2 \beta} \sum_r (t_r)^2 \biggr) \\
\noalign{\vskip 4mm}
&& \frac{1}{\sigma_{l}^{2}} = (\frac{2\pi l}{\beta})^2 +\frac{2}{\beta}\sum_{m}\sigma_{m}^{2} +\frac{3i}{\beta}\sum_{r}\frac {2\pi r}{\beta}
a_r +\frac{14}{\b}\sum_{m}\Delta_{m}^{2} + {2 \over \beta} \rho_0^2 \nonumber\\
\label{gap2}
&& \qquad - \frac{8}{\b}(\kl)^{2}\sigma_{l}^{2}\rho^{2}_{0}-\frac{14}{\b}\sum_{r+s+l=0}g_r g_s 
+\frac{1}{\b}\sum_{r+s+l=0}t_r t_s \\
\noalign{\vskip 4mm}
&& \frac{1}{a_r} = -i(\kr)^3 -\frac{3i}{\b}\kr\sum_{l}\sigma_{l}^{2} +\frac{4}{\b}\kr\sum_{s} \ks a_s
-\frac{5i}{2\b}\kr\rho_{0}^{2} \nonumber\\
&& \qquad - \frac{14i}{\b}\kr\sum_{l}\Delta_{l}^{2}
+\frac{8}{\b}(\kr)^4 a_r\rho_{0}^{2}
-\frac{14}{\b}\sum_{r+s+l=0}(\kl)^2\Delta_{l}^{2}g_s \nonumber\\
\label{gap3}
&& \qquad - \frac{14}{\b}\sum_{r+s+l=0}\epsilon_{l}^{2}g_s+\frac{1}{\b}\sum_{r+s+l=0}t_s u_l 
-\frac{1}{\b}\sum_{r+s+l=0} (\kl)^2 s_l t_s \\
\noalign{\vskip 4mm}
&& \frac{1}{\Delta_{l}^{2}} = (\frac{2\pi l}{\beta})^2+\frac{4}{\beta}\sum_{m}\sigma_{m}^{2} +\frac{4i}{\beta}\sum_{r}\frac {2\pi r}{\beta}a_r
+\frac{2}{\b}\rho_{0}^{2}+\frac{12}{\b}\sum_{m+n+l=0}\Delta_{m}^{2}\epsilon_{n}^{2}\nonumber\\
\label{gap4}
&& \qquad - \frac{12}{\b}\sum_{r+s+l=0}g_r g_s 
+\frac{4}{\b}(\kl)^2\sum_{r+s+l=0}a_r g_s-\frac{8}{\b}(\kl)^2\Delta_{l}^{2}\rho_{0}^{2} \\
\noalign{\vskip 4mm}
&& \frac{1}{g_r} = i \kr+\frac{12}{\b}\sum_{r+s+l=0}\Delta_{l}^{2} g_s -\frac{2}{\b}\sum_{r+s+l=0}
(\kl)^2\Delta_{l}^{2} a_s \nonumber\\
\label{gap5}
&&\qquad-\frac{2}{\b}\sum_{r+s+l=0}\epsilon_{l}^{2} a_s +\frac{4}{\b}\sum_{r+s+l=0}\sigma_{l}^{2} g_s 
+\frac{2}{\b}g_r \rho_{0}^{2} \\
\noalign{\vskip 4mm}
\label{gap6}
&& \frac{1}{\epsilon_{l}^{2}} = 1 +\frac{6}{\b}\sum_{m+n+l=0} \Delta_{m}^2 \Delta_{n}^2 +\frac{4}{\b}\sum_{r+s+l=0} a_r g_s \\
\noalign{\vskip 4mm}
\label{gap7}
&& \frac{1}{s_l} = -(\kl)^2-\frac{1}{\b}(\kl)^2\sum_{r+s+l=0} t_r a_s -\frac{2}{\b} (\kl)^2s_l \rho_{0}^{2} \\
\noalign{\vskip 4mm}
&& \frac{1}{t_r} = i\kr + \frac{1}{\b}\sum_{r+s+l=0} \sigma_{l}^{2} t_s -\frac{1}{2\b}\sum_{r+s+l=0}a_s u_l
+ {1 \over 2 \beta}\sum_{r+s+l=0}(\kl)^2 s_l a_s +\frac{1}{2\b}\rho_{0}^{2} t_r \nonumber\\
\label{gap8}
& & \qquad \\
\noalign{\vskip 4mm}
\label{gap9}
&& \frac{1}{u_l} = 1 +\frac{1}{\b}\sum_{r+s+l=0} t_r a_s
\eea

\subsection{Numerical methods}

The gap equations we have described form an infinite set of coupled
algebraic equations.  We now outline the numerical methods that we
used to solve these equations.  For additional details see appendix B
of \cite{KabatLifschytz}.

The first step is to reduce the infinite set of equations (\ref{gap1})
-- (\ref{gap9}) down to a finite set.  To do this we use the following
asymptotic forms of the propagators, which are valid at large
momenta.
\bea
\label{AsymptoticForms}
&& \sigma_l^2 \approx {1 \over (2 \pi l / \beta)^2 + m_\sigma^2} \qquad
a_r \approx {i \over 2 \pi r / \beta} \, {1 \over (2 \pi r / \beta)^2 + m_a^2} \\
\nonumber
&& \Delta_l^2 \approx {1 \over (2 \pi l / \beta)^2 + m_\Delta^2} \qquad
g_r \approx - {i 2 \pi r / \beta \over (2 \pi r / \beta)^2 + m_g^2} \qquad
\epsilon_l^2 \approx {(2 \pi l / \beta)^2 \over (2 \pi l / \beta)^2 + m_\epsilon^2} \\
\nonumber
&& s_l \approx - {1 \over (2 \pi l / \beta)^2 + m_s^2} \qquad
t_r \approx - {i 2 \pi r / \beta \over (2 \pi r / \beta)^2 + m_t^2} \qquad
u_l \approx {(2 \pi l / \beta)^2 \over (2 \pi l / \beta)^2 + m_u^2}
\eea
At leading order these are simply the tree-level
propagators.\footnote{This reflects the fact that the quantum
mechanics is free in the ultraviolet.}  Demanding that these
propagators satisfy the gap equations to the first subleading order of
an expansion in $1/({\rm momentum})$ fixes the asymptotic masses to be
\bea
\nonumber
m_\sigma^2 & = & {2 \over \beta} \sum_l \sigma_l^2 + {3 i \over \beta} \sum_r \kr a_r + {14 \over \beta} \sum_l \Delta_l^2 - {6 \over \beta} \rho_0^2 \\
\nonumber
m_a^2 & = & {3 \over \beta} \sum_l \sigma_l^2 + {4 i \over \beta} \sum_r \kr a_r + {14 \over \beta} \sum_l \Delta_l^2 - {11 \over 2 \beta} \rho_0^2 \\
\nonumber
m_\Delta^2 & = & {4 \over \beta} \sum_l \sigma_l^2 + {12 \over \beta} \sum_l \Delta_l^2 - {6 \over \beta} \rho_0^2 \\
\nonumber
m_g^2 & = & {4 \over \beta} \sum_l \sigma_l^2 + {12 \over \beta} \sum_l \Delta_l^2 - {2 \over \beta} \rho_0^2 \\
\nonumber
m_\epsilon^2 & = & - {4 i \over \beta} \sum_r \kr a_r + {12 \over \beta} \sum_l \Delta_l^2 \\
\nonumber
m_s^2 & = & - {i \over \beta} \sum_r \kr a_r - {2 \over \beta} \rho_0^2 \\
\label{masses}
m_t^2 & = & {1 \over \beta} \sum_l \sigma_l^2 - {1 \over 2 \beta} \rho_0^2 \\
\nonumber
m_u^2 & = & - {i \over \beta} \sum_r \kr a_r\,.
\eea

The next step is to fix a mode cutoff $N$.  For modes with $-N \leq
l,r \leq N$ we regard the Fourier modes of the propagators themselves as the
unknowns, while for modes outside this range we parameterize the
propagators in terms of the eight unknown asymptotic masses appearing
in (\ref{AsymptoticForms}).  The propagators with $-N \leq l,r \leq N$
are to be found by directly solving the relevant gap equations
(\ref{gap1}) -- (\ref{gap9}), while the asymptotic masses are to be
determined by solving the system of equations (\ref{masses}).  Note
that all these equations are coupled.  For example, we evaluate the
high-momentum parts of the loop sums that appear in (\ref{gap1}) --
(\ref{gap9}) and (\ref{masses}) analytically, in terms of the
asymptotic masses.

This leaves us with a finite set of equations.\footnote{$8N+13$
equations, to be precise, where we've taken advantage of the fact that
time-reversal invariance makes the (bosonic, fermionic) propagators
(even, odd) functions of their momenta.}  Our basic strategy is to
start at high temperatures $\beta \ll 1$, where we have the following
approximate solution to the gap equations.
\bea
\nonumber
\lambda & \approx & 0.418 \beta^{3/2} \qquad \sigma_0^2 \approx 0.209 \beta^{1/2}
\qquad \Delta_0^2 \approx 0.282 \beta^{1/2} \\
\epsilon_0^2 & \approx & 0.677 \qquad \qquad u_0 \approx 1
\eea
(All non-zero modes are approximately given by their free-field
values.)  Then we use the Newton-Raphson method
\cite{NumericalRecipes} to solve the system of equations at a sequence of
successively lower temperatures.  Our numerical results were obtained
starting at $\beta = 0.1$, with
\[
\beta \rightarrow {\rm min}(1.2 \, \beta,\, \beta+0.25)
\]
on each step, and with a mode cutoff $N = {\rm max}(3,\,5\beta)$.

Finally, after solving all the gap equations, we wish to compute the
free energy.  This has the expansion given in (\ref{betaFexpansion}),
which we truncate to
\be
\beta F \approx \beta F_0 + \langle S - S_0 \rangle_0 - {1 \over 2} \langle \left(S_{III}\right)^2 \rangle_{\C,0}\,.
\ee
That is, our approximation to $\beta F$ is simply the effective action
$I_{\rm eff}$ of (\ref{Ieff}).\footnote{The 0-brane action also has
6-point couplings, but it turns out that $\langle S_{VI}\rangle_0 = 0$ so these
terms do not contribute to our approximation for the free energy.}
The explicit expression for $I_{\rm eff}$ is given in appendix A.  To
calculate $I_{\rm eff}$ numerically we must make use of the asymptotic
forms (\ref{AsymptoticForms}).  For example we define the following
renormalized sum:
\bea
\nonumber
- \sum_l \log \sigma_l^2 & = & - \sum_{-N}^N \log \left[\sigma_l^2 \left((2 \pi l / \beta)^2 + m_\sigma^2\right)\right] \\
& & \qquad \quad + 2 \log (2 \sinh(\beta m_\sigma / 2))\,.
\eea

\section{Mean-field results for thermodynamic quantities}

In principle the trial action we have constructed contains a great
deal of information about correlation functions in the quantum
mechanics.  But in this section we will just present numerical results
for the behavior of three basic quantities: the free energy, the
Wilson loop, and the mean size of the state.

\subsection{Free energy}

At high temperatures, where the gauge theory is weakly coupled, we find
that the free energy of the system is
\be
\beta F = 6 \log \beta + {\cal O}(1)\,.
\ee
This result can be obtained analytically: the gap equations are
dominated by the bosonic zero modes, and the free energy is dominated
by $\beta F_0$.

In general, for a weakly-coupled theory in $0+1$ dimensions, one would
expect the free energy to behave like $\log \beta$.  But note that,
even though the gauge theory is weakly coupled at high temperature,
the perturbation series is afflicted with IR divergences.  Thus, to
determine the coefficient of the logarithm (which depends on the value
of the dynamically generated IR cutoff) one must re-sum part of the
perturbation series.  This is a well-known phenomenon in finite
temperature field theory \cite{DolanJackiw}.  In any case, we expect
{\em a priori} that mean-field methods give good results in
the high temperature regime.

\begin{figure}
\epsfig{file=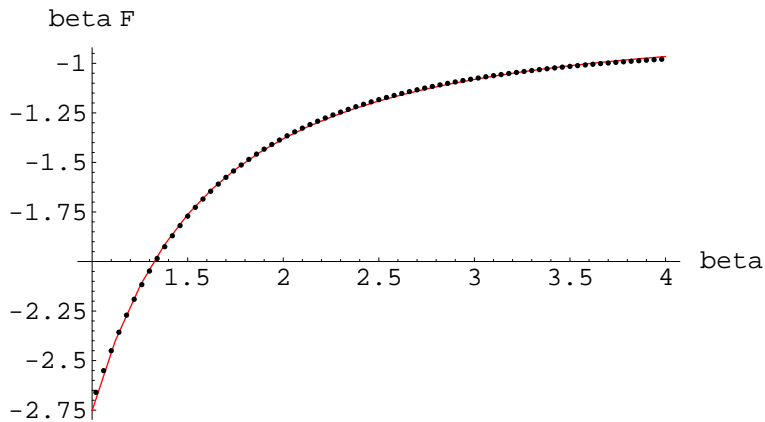}
\caption{The solid curve is the power law fit (\ref{results:fit})
for $\beta F$.  The data points are calculated from numerical
solutions to the gap equations.}
\end{figure}

As the temperature is lowered the behavior of the free energy changes:
at $\beta \approx 0.7$ we find that it begins to roll over and fall
off as a non-trivial power of the temperature.  In the range $1 <
\beta < 4$ the numerical results for the free energy are well fit by
\be
\label{results:fit}
\beta F \approx - 0.79 - 2.0 ~ \beta^{-1.7}\,.
\ee
This fit to the numerical results is illustrated in Fig.~1.  Note that
supersymmetry is crucial in making such power-law behavior possible.
Without supersymmetry the free energy would behave as $\beta F \approx
\beta E_0$ in the low temperature regime ($\beta > 1$), where $E_0$ is
the ground state energy of the system.

We obtained (\ref{results:fit}) by performing a Levenberg-Marquardt
nonlinear least-squares fit \cite{NumericalRecipes} to 75 numerical
calculations of the free energy, carried out in the temperature range
$1 \leq \beta \leq 4$. To estimate the uncertainty in the best fit
parameters we varied the window of $\beta$ over which the fit was
performed (fitting over the ranges $2<\beta<4$ and $1<\beta<3$), which
leads to: $-0.79 \pm 0.06$, $-2.0 \pm 0.1$ and $-1.7\pm 0.2$.

It is quite remarkable that the power law (\ref{results:fit}) is in
excellent agreement with the semiclassical black hole prediction
\cite{imsy,KlebanovTseytlin}
\begin{equation}
\beta F = - 4.12 ~ \beta^{-1.80}\,.
\end{equation}
The exponents differ by 6\% while the coefficients of the power-law
differ by a factor of 2.  (An additive constant appears in the
mean-field approximation for the free energy.  We will generally
ignore this `ground state degeneracy', since it seems to be an
artifact of the approximation when applied to systems with a
continuous spectrum.  Similar behavior was noted in
\cite{KabatLifschytz}.)  In a toy model studied in
\cite{KabatLifschytz} it was noted that higher order terms in the
expansion of the free energy (\ref{betaFexpansion}) appear with
approximately the same power law dependence on temperature as the
leading term.  Thus by computing higher-order corrections one might
hope for better agreement of the overall coefficient, with the power
law essentially unchanged.

As we go to still lower temperatures, we find that the energy
$\partial (\beta F)/\partial\beta$ calculated in the mean-field
approximation begins to drop below the energy of the black hole.  In
fact the mean-field energy becomes negative around $\beta = 5.8$.
Ultimately, as $\beta \rightarrow \infty$, the mean-field energy does
asymptote to zero, as required by the ${\cal N} = 2$ supersymmetry
which is manifest in the approximation.  But a negative energy clearly
reflects some problem with the approximation.

Fortunately, we can be rather precise about exactly where the
approximation is going wrong: the difficulty is with the
Schwinger-Dyson gap equation we have been using to fix the value of
the one-plaquette coupling $\lambda$.  Although we do not know how to
write down a better gap equation for $\lambda$, we can give a {\em
prescription} for fixing $\lambda$, that will allow us to obtain
reasonable results at much lower values of the temperature. This may
be regarded either as a check on our understanding of why the
approximation is breaking down, or as a way of building a model for
the black hole that can be used at lower temperatures.  Our
prescription for fixing $\lambda$ is simply that, when $\beta > 2.5$
(the midpoint of our range $1 \le \beta \le 4$), we choose $\lambda$
so that the free energy is given by (\ref{results:fit}).  The energy
$E = \partial (\beta F) / \partial \beta$ calculated with this
prescription is shown in Figs.~2 and 3.

\begin{figure}
\epsfig{file=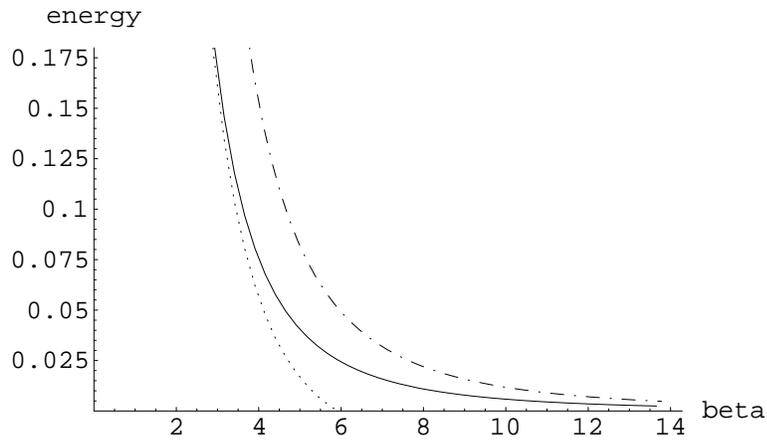}
\caption{Energy vs.~$\beta$.  For $\beta >2.5$ fixing $\lambda$ by
fitting $\beta F$ to a power law leads to the solid middle line, while
the Schwinger-Dyson gap equation for lambda leads to the lower dashed 
line.  The upper dot-dashed line is the semiclassical energy of the
black hole.}
\end{figure}

\begin{figure}
\epsfig{file=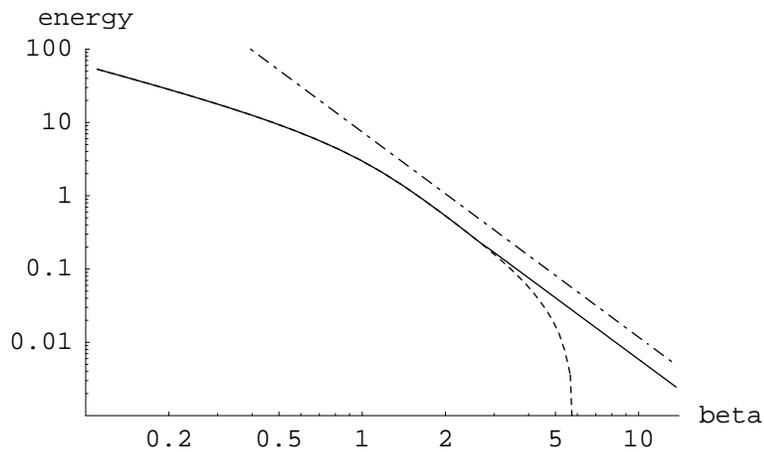}
\caption{Energy vs.~$\beta$.  Same as Fig.~2, but plotted on a
$\log$--$\log$ scale.}
\end{figure}

\subsection{Wilson loop}

In our approximation the expectation value of the timelike Wilson loop
$U = P e^{i \oint d\tau A_0}$ is controlled by the one-plaquette
coupling $\lambda$, as in (\ref{WilsonVEV}).  A key feature of the
one-plaquette action is that a large-$N$ phase transition occurs at
$\lambda = 2$ \cite{GrossWitten}.  At this value of the coupling the
eigenvalues of $U$ spread out around a circle, and become sensitive to
the fact that the gauge field is a periodic variable.  It has been argued that
just such a phase transition is expected to occur in 0-brane quantum
mechanics, as the system moves from weak coupling into the
supergravity regime \cite{KabatLifschytz}.

Our mean-field results for $\lambda$ are shown in Fig.~4.  We present
the results for $\lambda$ that are obtained by solving the
Schwinger-Dyson equation (\ref{AooSD}), as well as the results that
are obtained from our prescription of fitting $\beta F$ to a power
law.

\begin{figure}
\epsfig{file=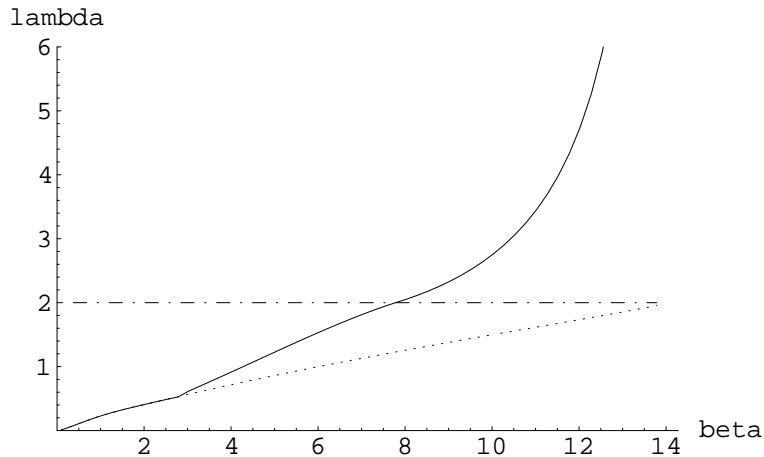}
\caption{The one-plaquette coupling $\lambda$ vs.~$\beta$.  The
Gross-Witten transition occurs when $\lambda = 2$.  For $\beta < 2.5$ we
use the Schwinger-Dyson gap equation to determine $\lambda$.  For
$\beta > 2.5$ the Schwinger-Dyson gap equation gives the dashed
line, while fitting $\beta F$ to a power law gives the solid
line.}
\end{figure}

Note that in both cases, $\lambda$ increases monotonically with
$\beta$.  The Gross-Witten phase transition takes place when $\lambda
= 2$; with the prescription of fitting $\beta F$ to a power law this
value is reached at $\beta = 7.8$.  Thus, as expected, a phase
transition takes place as the system moves into the supergravity
regime \cite{KabatLifschytz}.  By adopting the prescription of fitting
$\beta F$ to a power law, we cannot say anything about the order of
the phase transition.  But if one takes the Schwinger-Dyson result for
$\lambda$ seriously, then the Gross-Witten transition occurs at $\beta
= 14.2$, and is weakly second order (the second derivative of the free
energy drops by $0.01$ in crossing the transition).

Our prescription for choosing $\lambda$ by fitting $\beta F$ to a
power law begins to break down around $\beta = 14$, as we find that
$\lambda$ rapidly diverges as $\beta$ approaches 14.\footnote{The
Schwinger-Dyson gap equation for $\lambda$ has solutions at all
temperatures.}  By itself, this is not necessarily a problem: infinite
$\lambda$ simply means that the Wilson loop is uniformly distributed
over $U(N)$.  But unfortunately, we do not have a good prescription
for continuing past this temperature.  Evidently some of the other gap
equations (not just the gap equation for $\lambda$) start to break
down at this point.  Note that this breakdown does not occur until
well into the strong coupling regime, as an inverse temperature $\beta
= 14$ corresponds to an effective gauge coupling $g_{\rm eff}^2 =
\beta^3 \approx 3 \times 10^3$.

\subsection{Mean size}

Finally, let us comment on the average `size' of the state.  In our
approximation the scalar fields $X^i(\tau)$ and $\phi^a(\tau)$ are
Gaussian random matrices, and their eigenvalues obey a Wigner
semi-circle distribution.  We can define the size of the state in terms
of the quantities
\bea
\label{radius}
R^2_{\rm gauge}  &=&  {1 \over N} \langle {\rm Tr}
\left(X^i(\tau)\right)^2 \rangle_0 ~, \nonumber \\
R^2_{\rm scalar} & =&  {1 \over N} \langle
{\rm Tr} \left(\phi^a(\tau)\right)^2 \rangle_0 \,.
\eea
The radius of the Wigner semi-circle, given by $2 \sqrt{R^2}$, is shown
in Fig.~5.  Note that the radius stays fairly constant in the region
corresponding to the black hole.  However, because the superfield
formalism we are using does not respect the full $SO(9)$ invariance,
the radius measured in the scalar multiplet directions is not the same
as the radius measured in the gauge multiplet directions.  At $\beta = 14$
we find
\[
2 R_{\rm scalar} = 1.81 \qquad\quad 2 R_{\rm gauge} = 0.80\,.
\]
This shows that, as expected, the trial action does not respect the
underlying $SO(9)$ invariance.  Nonetheless, the trial action may
provide a useful approximate description of the black hole density
matrix in the supergravity regime.

\begin{figure}[ht]
\epsfig{file=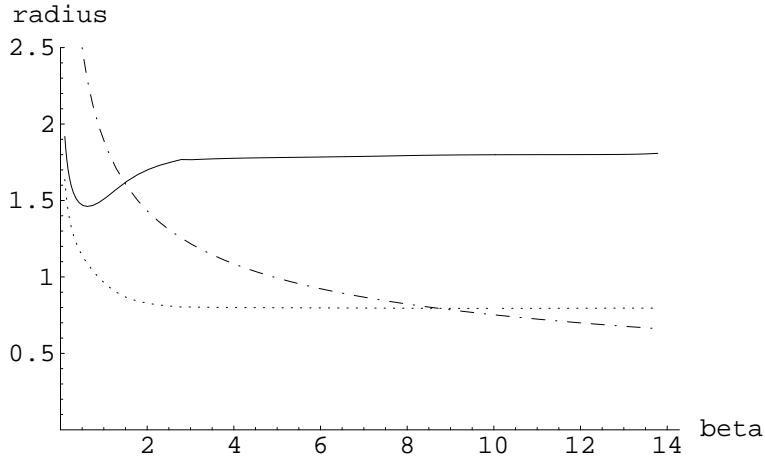}
\caption{Range of eigenvalues (radius of the Wigner semi-circle)
vs.~$\beta$.  The upper solid curve is for the scalar fields in the
scalar multiplets; the lower dotted curve is for the scalar fields in
the gauge multiplet.  The dashed curve is the Schwarzschild
radius of the black hole.  These results were calculated with $\beta
F$ fit to a power law for $\beta >2.5$.}
\end{figure}

In Fig.~5 we have also plotted the Schwarzschild radius of the black
hole
\be
\label{SchwarzschildRadius}
U_0/2\pi = 1.89~\beta^{-2/5}\,.
\ee
Note that, as the temperature decreases, the Schwarzschild radius
becomes much smaller than the radius of the eigenvalue distributions.
It seems appropriate to identify the radius of the eigenvalue
distributions with the size of the region $U \ll (g^2_{\rm YM}
N)^{1/3}$ in which 10-dimensional supergravity is valid \cite{imsy}.

This brings up a subtle issue.  The Higgs fields of the gauge theory
are expected to correspond to spatial coordinates in the supergravity
geometry (\ref{metric}).  But one is always free to reparameterize the
radial coordinate in supergravity.  In (\ref{SchwarzschildRadius}) we
have implicitly made use of the naive identification $X = U / 2 \pi$,
where $X$ is a Higgs field and $U$ is the supergravity coordinate
appearing in (\ref{metric}).  This can be justified at zero
temperature, because supersymmetry fixes the mass of a BPS stretched
string in the gauge theory to be given by the tree-level formula $m_W
= X$, while in supergravity one has $m_W = U / 2 \pi$ \cite{imsy}.
However this particular identification is not appropriate at finite
temperature.  A proposal for relating the two radial coordinates has
been presented in \cite{radial}.

An unambiguous way to fix the relation is to use the fact that $m_W =
(U-U_0)/2\pi$ in the non-extremal black hole geometry.  By computing
$m_W$ in the gauge theory, the mapping between the Higgs field $X$ and
the supergravity coordinate $U$ can be fixed. However one must first
take account of the fact that one has a continuous distribution of
masses in the quantum mechanics. The spacetime geometry will only
correspond to the lightest of these states as we discuss in section 5.
This procedure will be studied further in \cite{IKLL}.

\section{Propagators and spectral weights}

Important information about spacetime geometry is encoded in the
spectrum of single-string excitations in the quantum mechanics.  To
extract this information from the Euclidean propagators we introduce a
spectral representation for the 2-point functions.  By inserting
complete sets of states, one can show that at finite temperature the
analog of the Lehmann spectral representation takes the form
\be
\label{SpectralRep}
\langle \phi(\tau) \phi(0) \rangle = \int_0^\infty d\omega \, \rho(\omega) \, {\cosh \omega(\tau - \beta/2)
\over 2 \omega \sinh (\beta \omega / 2)} \qquad 0 \le \tau \le \beta
\ee
where the spectral weight $\rho(\omega)$ is defined as the thermal average
\[
\rho(\omega) = {1 \over Z} \sum_m e^{-\beta E_m} \sum_{n > m} \vert
\langle n
\vert \phi \vert m\rangle \vert^2 \, 2\omega \left(
1 - e^{-\beta \omega}\right) \delta(\omega-E_n+E_m)\,.
\]
We can interpret $\rho(\omega) d\omega$ as the effective number of
single-string microstates with a mass between $\omega$ and
$\omega + d\omega$.  We will apply this to the scalar fields in the
scalar multiplets, setting
\[
\Delta_l^2  =  \int_0^{\infty} d \omega \rho(\omega) {1 \over
(2 \pi l / \beta)^2 + \omega^2}\,.
\]

\begin{figure}
\epsfig{file=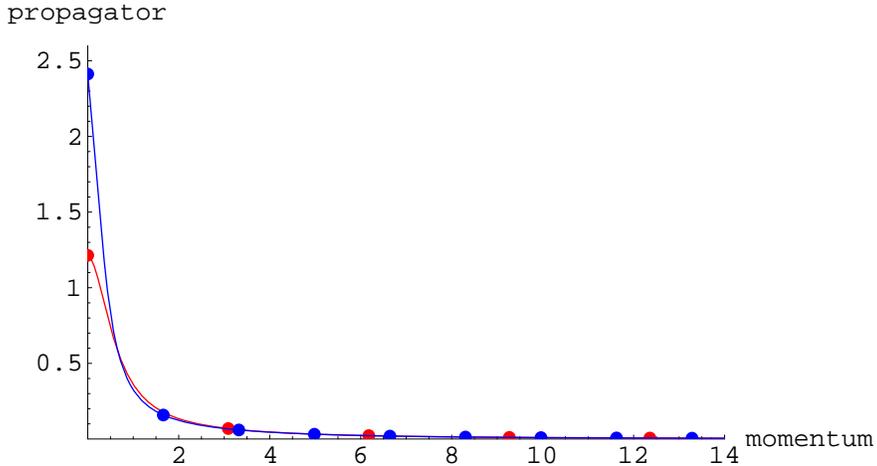}
\caption{The 2-point function of the scalar fields in the scalar
multiplets, calculated at Matsubara frequencies. The solid curve is
a fit to the propagator with a twin peak ansatz for the density of
states. The upper blue curve is for $\beta=3.78$, the lower red curve for
$\beta=2.03$.}
\end{figure}

\begin{figure}
\epsfig{file=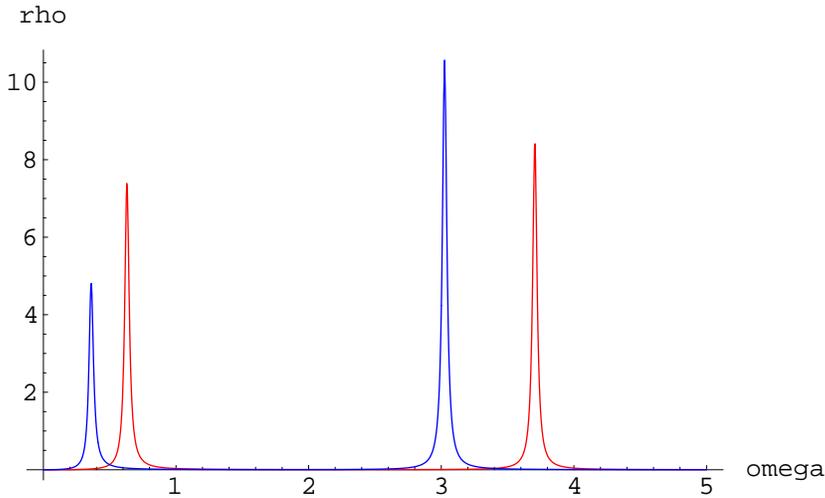}
\caption{The effective density of states arising from the scalar
multiplet. We show the result for two different temperatures: the
blue peaks are for $\beta=3.78$, the red for $\beta=2.03$.}
\end{figure}

In general, solving the inverse problem to extract the density of
states from $\Delta_l^2$ is a difficult numerical problem, which we
analyze in detail in \cite{IKLL}.  However some gross features of the
spectral density can be easily seen.  Consider the $\Delta^2$
propagator which is plotted in Fig.~6.  At large frequency the
behavior of the propagator is controlled by the asymptotic mass
$m_\Delta^2$ (\ref{masses}).  This mass is of order one in 't Hooft
units; for example $m_\Delta = 2.8$ at $\beta = 3.78$.  This indicates
that states with a mass of order one are present in the spectrum.  In
addition, note that a clear enhancement of the propagator at small
frequency compared to its asymptotic form can be seen in Fig.~6.  This
suggests the presence of light states in the spectrum, with a mass of
order the temperature.  These light states make the dominant
contribution to the entropy.

To express this in a more quantitative way, we will make an ansatz for
the form of $\rho(\omega)$.  The ansatz will allow us to estimate the spectral
density without performing a complete analysis of the inverse problem.
Our ansatz is motivated by the results of \cite{IKLL}, where we
considered a 0-brane probe of the black hole background.  The
full analysis of \cite{IKLL} shows that the density of states consists of two
narrow peaks (i.e. narrower than a scale of order the temperature),
one centered at frequency of order the temperature, the other at
frequency of order the 't Hooft coupling. Motivated by this result, we
introduce the following ansatz
\be
\label{denstat}
\rho(\omega) = { \omega z_1\over {(\omega-m_1)^2+\gamma^2}} +{
  \omega z_2\over {(\omega-m_2)^2+\gamma^2}} + (\omega \to -\omega)\,,
\ee
where we fix the width $\gamma=0.02$ (any number much smaller than $T$
will do).  The parameters in (\ref{denstat}) are determined by fitting
to the $\Delta_l^2$ propagator.  Note that the small value of $\gamma$
is motivated by \cite{IKLL}; we could not hope to extract such a small
width by fitting the ansatz just at Matsubara frequencies.  The
resulting densities of states are shown in Fig.~7. As can be seen from
the plot, we obtain a very good fit to the propagators at the
Matsubara frequencies.  The fit determines the parameters $z_i$ and
$m_i$ with uncertainties at the level of a few percent.

This result clearly indicates the presence of both the high and low
frequency states mentioned above.  It also points to a clear
separation between the two sets of states -- a separation which will
play an important role in the next section.

\section{Resolving spacetime geometry}

In our mean-field approximation, we have modelled the cloud of 0-branes that
make up the black hole using Gaussian random matrices.  As we
discussed in section 3.3, the eigenvalues of these matrices have
very large quantum fluctuations.  Within the gauge theory, we find that the
scale of these fluctuations is set by the 't Hooft coupling.
\be
\label{NaiveSize}
R^2 = {1 \over N} \langle {\rm Tr} \phi(\tau)^2 \rangle_0 \, \sim \, \left(g^2_{YM} N\right)^{2/3}
\ee
In terms of supergravity, this means that the positions of the
0-branes that make up the black hole have very large quantum
fluctuations.  Indeed they fluctuate over roughly the entire region
(of size $U \sim \left(g^2_{YM} N\right)^{1/3}$ \cite{imsy}) in which
supergravity is valid.  One might suspect that these large
fluctuations are an artifact of our approximation, but it has been
argued that the scaling (\ref{NaiveSize}) is an intrinsic feature of
0-brane quantum mechanics \cite{Polchinski}.

This raises a very interesting question.  How can we recover local
spacetime physics from the quantum mechanics?  In particular, given
the large fluctuations (\ref{NaiveSize}), how can we resolve the
horizon of the black hole?

The answer is that local spacetime physics only arises as a {\em low
energy} approximation to the quantum mechanics.  To recover local
spacetime physics from the quantum mechanics we must introduce a
resolving time, and integrate out high frequency degrees of
freedom\footnote{We are grateful to Leonard Susskind and Emil Martinec
for discussions on this topic.}.  The point is that most of the $N^2$
degrees of freedom in the quantum mechanics have a very large
frequency, set by the 't Hooft coupling.
\[
\omega \sim \left(g^2_{YM} N\right)^{1/3}
\]
{}From the supergravity point of view, this energy scale corresponds to the
energy of a string that stretches across the entire region in which
supergravity is valid.  A low energy observer within supergravity
cannot resolve such high-frequency fluctuations.  Therefore, to
recover local spacetime physics from the quantum mechanics, we must
first introduce a resolving time $\epsilon$, and integrate out all
modes with frequencies larger than $1 / \epsilon$.  With an
appropriate choice of resolving time, we should recover the expected
result, that the 0-branes only fluctuate over a region whose size is set
by the horizon of the black hole.

We begin by discussing a single harmonic oscillator.  At finite
temperature the fluctuation in the oscillator position coordinate is
\[
\langle x^2\rangle = {1 \over 2 \omega \tanh (\beta \omega / 2)}\,.
\]
We introduce a resolving time, by smearing the Heisenberg picture operators over
a Lorentzian time interval $\epsilon$.
\[
\bar{x} = \int_{-\infty}^\infty {dt \over \epsilon \sqrt{\pi}} \, e^{-t^2 / \epsilon^2} x(t)
\]
The fluctuations in the smeared operators are suppressed when $\omega > 1/\epsilon$.
\[
\langle \bar{x}^2\rangle = {e^{-\omega^2 \epsilon^2 / 2} \over 2 \omega \tanh (\beta \omega / 2)}
\]
To take this over into 0-brane quantum mechanics, we
use the spectral representation (\ref{SpectralRep}).
The fluctuations in the field are given by
\be
\label{phi2SpecRep}
\langle \phi^2\rangle = \int_0^\infty d\omega \, \rho(\omega) \, {1 \over 2 \omega \tanh(\beta \omega / 2)}\,.
\ee
Following our treatment of the harmonic oscillator, we can introduce a resolving time by setting
\be
\label{resolving1}
\langle \bar{\phi}^2\rangle = \int_0^\infty d\omega \, \rho(\omega) \, {e^{-\omega^2 \epsilon^2 / 2} \over 2 \omega \tanh(\beta \omega / 2)}\,.
\ee

One might worry that this definition of horizon radius is ambiguous,
since it seems to depend on the choice of resolving time.
Fortunately, from Fig.~7, we see that the spectral density consists of
two well-separated peaks.  The low-frequency peak corresponds to
states with an energy of order the temperature; these states can be
thermally excited and should be included in the fluctuations which
make up the horizon.  The high-frequency peak corresponds to states
with an energy of order the 't Hooft coupling, states which should be
integrated out to see agreement with supergravity.  Thus any resolving
time which keeps the low-frequency peak and integrates out the
high-frequency peak is acceptable, and will produce the same horizon
radius.

\begin{figure}
\epsfig{file=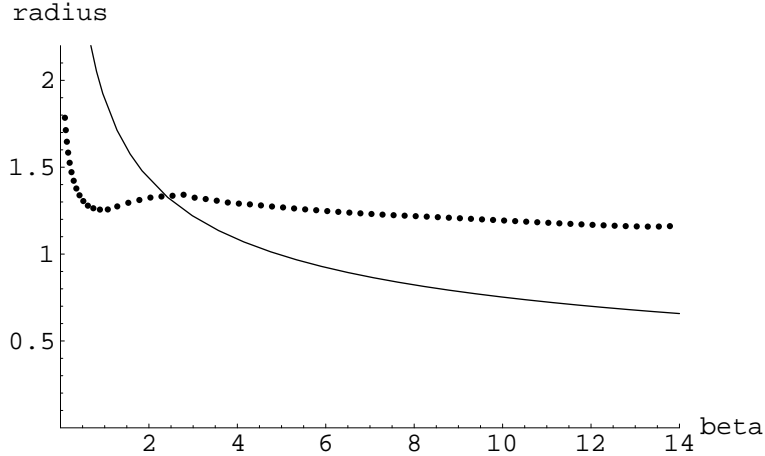}
\caption{The dotted curve is the smeared radius of the Wigner
semi-circle.  The solid curve is the Schwarzschild radius of
the black hole, as measured in the $U$ coordinate.}
\end{figure}

For reasonable values of $\epsilon$ we can easily estimate $\langle
\bar{\phi}^2\rangle$.  Rather than use the Gaussian cutoff
(\ref{resolving1}), it turns out to be more convenient to define a
time-averaged size by introducing a factor $1/\cosh(\beta \omega/2)$
into the integrand of (\ref{phi2SpecRep}).
\be
\label{resolving2}
\langle \bar{\phi}^2\rangle \equiv \int_0^\infty d\omega \, \rho(\omega) \, {1 \over 2 \omega \sinh(\beta \omega / 2)}
\ee
The extra factor has the effect of cutting off the integral at $\omega
\approx 1/\beta$.  This corresponds to a reasonable choice of resolving time,
$\epsilon \approx \beta$. Defined in this way, our estimate for the
time-averaged fluctuations in the 0-brane positions is simply given by
a Euclidean Green's function, {\it c.f.}~(\ref{SpectralRep}).
\[
\langle \bar{\phi}^2\rangle \,\, = \,\, \langle \phi(\beta/2) \, \phi(0)\rangle
\]
This is easily calculated as a Fourier transform of our momentum-space
propagators.  In Figure 6 we plot the resulting smeared radius of the
Wigner semicircle
\[
\bar{R} = 2 \sqrt{(7 \bar{R}_{\rm scalar}^2 + 2 \bar{R}_{\rm gauge}^2)/9}
\]
as a function of $\beta$ (we average over the scalar fields in the
gauge and scalar multiplets).\footnote{The result for $\bar{R}$ is
dominated by the contribution of the zero-frequency Matsubara mode.}
The time-averaged fluctuations in the 0-brane positions go down with
temperature.  This is the expected behavior for the size of these
black holes.  The result for $\bar{R}$ is in rough agreement with the
Schwarzschild radius $U_0$ of the black hole (\ref{SchwarzschildRadius}),
which we show in the same plot.

\section{Conclusions and future directions}

In this paper we have developed an approximation scheme for 0-brane
quantum mechanics at strong coupling and finite temperature.  We
presented an ansatz for a trial action which captures some of the
behavior of the large-$N$ quantum mechanics.  The parameters appearing
in the trial action are chosen according to a set of gap equations
which resum an infinite set of planar diagrams.  The approximation
automatically respects 't Hooft large-$N$ counting, and also partially
respects the supersymmetries and R-symmetries of the quantum
mechanics.

Our main result is that we find good agreement with black hole
thermodynamics over the temperature range $1 < \beta < 4$.  In
addition, we studied the behavior of a Wilson loop, and found that as
expected a large-$N$ phase transition occurs as the system enters the
supergravity regime.  We also presented results on the mean size of
the system, and argued that in the supergravity regime this mean size
(as measured by quantum fluctuations) exceeds the Schwarzschild radius
of the dual black hole.  We could nonetheless recover the
Schwarzschild radius from the quantum mechanics, by introducing a
suitable resolving time; to make our prescription unambiguous it was
important that the spectral density showed a clear separation between
light and heavy degrees of freedom.

We would like to emphasize that in the temperature range $1 < \beta <
4$ our approximation applies strictly to the gauge theory, and makes
no use of supergravity information.  We presented a prescription for
fixing the value of the Wilson loop, which allowed us to extend the
agreement up to $\beta = 14$.  The prescription, however, relies on
supergravity inputs.

Our results are based on several technical developments in the use of
mean-field methods.  Some of these developments were reported in our
previous work \cite{KabatLifschytz}.  In the present paper, the main new
technical problem we faced was the difficulty of treating gauge
theories using mean-field methods.  By working in the gauge
(\ref{GaugeChoice}), many of the Slavnov-Taylor identities become trivially
satisfied.  This gauge choice was instrumental in enabling us to find
a consistent set of gap equations, that could be solved in the
strong-coupling regime.

There are two perspectives that one could take on this subject.  The
`supergravity' perspective is that, since gauge theory is better
understood than quantum gravity, we should try to study supergravity
phenomena from the gauge theory point of view.  The `field theory'
perspective is to regard 0-brane quantum mechanics as an interesting
laboratory for developing and testing methods to study field theories
at strong coupling.

Depending on which perspective one adopts, there are several
interesting possible directions for future work.  From the field
theory point of view, it would be interesting to apply mean-field
methods to other models, to better understand the range of validity of
these techniques.  Let us mention one possibility.  One can apply our
techniques to pure ${\cal N} = 2$ gauge theory, simply by dropping the
scalar multiplets.  The resulting system of gap equations does not
have a solution in the low temperature regime $\beta > 1$.  Presumably
this can be related to the fact that the pure gauge model breaks
supersymmetry spontaneously \cite{Witten}.  It would be interesting to
understand this connection in more detail.

It would also be interesting to have additional tests of our
approximation scheme.  The supergravity makes further predictions for
the behavior of the gauge theory, which could be tested.  For example,
in \cite{yoneya} a set of predictions were made for the scaling
exponents of two-point functions of certain operators at zero
temperature. These were extracted by computing Green's functions in
the extremal supergravity background, and taking a large time/low
frequency limit. It was argued that these predictions follow from a
generalized conformal symmetry that appears in the 't Hooft limit
\cite{jevicki}.

It would be interesting to test these predictions against our
numerical results. Even at finite temperature, one would still expect
to recover the scaling behavior for frequencies satisfying $T \ll
\omega \ll (g^2_{YM} N)^{1/3}$. Unfortunately, the correlators which
are predicted to have a scaling behavior involve composite operators
whose two-point functions we have not yet computed. We hope to study
this question further in the future.

Another interesting direction would be to better understand the
duality between gravity and gauge theory.  In particular, it would be
interesting to understand better how the supergravity properties of
spacetime locality and causality emerge from the gauge theory.  One
might hope to see that the horizon of the black hole is reflected in
the dynamics of the gauge theory along the lines of \cite{causality}.
Also, as we mentioned at in section 3.3, there is the subtle question
of which radial coordinate in supergravity corresponds to the gauge
theory Higgs fields.  To address these sorts of issues, it is natural
to introduce a 0-brane to probe the supergravity background.  The
probe has a dual description in terms of a spontaneously broken gauge
theory.  In \cite{IKLL}, we use mean-field methods to study this
problem.

Ultimately, one might hope to use mean-field methods to study
non-equilibrium processes in the gauge theory at strong coupling,
perhaps using some sort of thermofield formalism \cite{Das}.  For
example, it would be extremely interesting to study scattering of a
graviton wavepacket off a black hole.  Could one see correlations in
the outgoing Hawking radiation?

\bigskip
\centerline{\bf Acknowledgements}
We are grateful to Norihiro Iizuka, Anton Kapustin, Emil Martinec,
Vipul Periwal, Martin Schaden and Mark Stern for valuable discussions.
DK wishes to thank New York University for hospitality.  The work of
DK is supported by the DOE under contract DE-FG02-92ER40699.  GL would
like to thank the Aspen Center for Physics and Tel-Aviv University
for hospitality.  The work
of GL is supported in part by the NSF under grant PHY-98-02484 and in 
part by US Israel bi-national science foundation grant 96-00294.
 DL wishes to
thank the Aspen Center for Physics, the Abdus Salam International
Centre for Theoretical Physics and the Erwin Schroedinger Institute
program in Duality, String Theory and M-Theory for hospitality during
the course of this research.  The research of DL is supported in part
by DOE grant DE-FE0291ER40688-Task A.

\appendix
\section{2PI effective action}

The two-loop, 2PI effective action is defined by
\be
I_{\rm eff} = \beta F_0 + \langle S_{II} + S_{IV} - S_0\rangle_0 - \frac{1}{2} \langle\left(S_{III}\right)^2\rangle_{\C,0}\,.
\ee
We adopt units which effectively set $g^2_{YM} N = 1$, and suppress
the overall factor of $N^2$ in the free energy.  Then the free energy
of the trial action is given by
\bea
\nonumber
\beta F_0 & = & \beta F_\scriptlap(\lambda) - \sum_l \log \sigma_l^2 + \sum_r \log a_r \\
\nonumber
& & - {7 \over 2} \sum_l \log \Delta_l^2 + 7 \sum_r \log g_r - {7 \over 2} \sum_l \log \epsilon_l^2 \\
& & + \sum_{l \not= 0} \log s_l - 2 \sum_r \log t_r + \sum_l \log u_l
\eea
where the free energy of the one-plaquette model is
\be
\beta F_\scriptlap(\lambda) = \left\lbrace
\begin{array}{ll}
-{2 \over \lambda} - {1 \over 2} \log {\lambda \over 2} + {3 \over 4} & \quad \lambda \leq 2 \\
\noalign{\vskip 2mm}
- 1 / \lambda^2 & \quad \lambda \geq 2\,.
\end{array} \right.
\ee
We also have
\bea
\nonumber
& & \langle S_{II} - S_0\rangle_0 = {N \over \lambda} \langle {\rm Tr} (U + U^\dagger)\rangle_\scriptlap \\
\nonumber
& & \qquad + \sum_l \left( (\kl)^2 \sigma_l^2 - 1 \right) 
+ \sum_r \left( i (\kr)^3 a_r + 1 \right) \\
\nonumber
& & \qquad + {7 \over 2} \sum_l \left( (\kl)^2 \Delta_l^2 - 1 \right)
+ 7 \sum_r \left( -i \kr g_r + 1 \right)
+ {7 \over 2} \sum_l \left( \epsilon_l^2 - 1 \right) \\
& & \qquad + \sum_{l\not=0} \left(({2 \pi l \over \beta})^2 s_l + 1\right)
+ 2 \sum_r \left(i {2 \pi r \over \beta} t_r - 1\right) - \sum_l \left(u_l - 1\right)
\eea
where the one-plaquette model contribution is
\be
{N \over \lambda} \langle {\rm Tr} (U + U^\dagger)\rangle_\scriptlap = \left\lbrace
\begin{array}{ll}
{2 \over \lambda} - {1 \over 2} & \quad \lambda \leq 2 \\
\noalign{\vskip 2mm}
2 / \lambda^2 & \quad \lambda \geq 2\,.
\end{array} \right.
\ee
We also have the contribution of the 4-point couplings,
\bea
\nonumber
& & \langle S_{IV}\rangle_0 = - {2 \over \beta} \sum_{r,s} \kr \ks a_r a_s + {3 i \over \beta} \sum_{l,r} \kr a_r \sigma_l^2
+ {1 \over \beta} \sum_{l,m} \sigma_l^2 \sigma_m^2 \\
\nonumber
& & \qquad \qquad + {14 \over \beta} \sum_{l,m} \Delta_l^2 \sigma_m^2 + {14 i \over \beta} \sum_{l,r} \Delta_l^2 \kr a_r
+ {7 \over \beta} \sum_l \Delta_l^2 \rho_0^2 \\
& & \qquad \qquad + {5 i \over 2 \beta} \sum_r \kr a_r \rho_0^2 + {2 \over \beta} \sum_l \sigma_l^2 \rho_0^2
\eea
where the two-point function of $A_{00}$ is defined in (\ref{rho02}).  Finally, the contribution of the three-point
couplings is given by
\bea
\nonumber
& & - {1 \over 2} \langle \left(S_{III}\right)^2\rangle_{\C,0} = - {4 \over \beta} \sum_l (\kl)^2 (\sigma_l^2)^2 \rho_0^2
- {4 \over \beta} \sum_r (\kr)^4 (a_r)^2 \rho_0^2 \\
\nonumber
& & \qquad\qquad + {14 \over \beta} \sum_{l+r+s=0} (\kl)^2 \Delta_l^2 a_r g_s + {14 \over \beta} \sum_{l+r+s=0} \epsilon_l^2 a_r g_s
- {14 \over \beta} \sum_{l+r+s = 0} \sigma_l^2 g_r g_s \\
\nonumber
& & \qquad\qquad - {14 \over \beta} \sum_l (\kl)^2 (\Delta_l^2)^2 \rho_0^2 - {7 \over \beta} \sum_r (g_r)^2 \rho_0^2 \\
\nonumber
& & \qquad\qquad + {21 \over \beta} \sum_{l+m+n = 0} \Delta_l^2 \Delta_m^2 \epsilon_n^2 - {42 \over \beta} \sum_{l+r+s = 0} \Delta_l^2 g_r g_s \\
\nonumber
& & \qquad\qquad + {1 \over \beta} \sum_{l+r+s = 0} \left(\sigma_l^2 t_r t_s - u_l t_r a_s + (\kl)^2 s_l t_r a_s \right) \\
& & \qquad\qquad + {1 \over 2 \beta} \sum_r (t_r)^2 \rho_0^2 + {1 \over \beta} \sum_l (\kl)^2 (s_l)^2 \rho_0^2\,.
\eea


\end{document}